\documentstyle[preprint,aps]{revtex}

\newcommand{\disregard}[1]{}

\tighten
\begin{document}


\title{
 Quadrupole and
Hexadecapole Correlations in Rotating
Nuclei Studied within the Single-$j$ Shell Model
}

\author  {P. Magierski$^a$,
                K. Burzy\'nski$^b$,
E. Perli\'nska$^b$, J. Dobaczewski$^b$,
                and W. Nazarewicz$^{a-d}$\\[5mm]
$^a${\em Institute of Physics, Warsaw University of Technology}\\
{\em ul. Koszykowa 75, PL--00-662 Warsaw, Poland}\\[2mm]
            $^b${\em Institute of Theoretical Physics,
                Warsaw University}\\
    {\em Ho\.za 69, PL-00-681 Warsaw, Poland}\\[2mm]
$^c${\em Department of Physics and Astronomy, University of  Tennessee}\\
       {\em Knoxville, Tennessee 37996, U.S.A.}\\[2mm]
$^d${\em Physics Division, Oak Ridge National Laboratory}\\
       {\em P. O. Box 2008, Oak Ridge, Tennessee 37831, U.S.A.}
}

\maketitle

\begin{abstract}
The influence  of  quadrupole
and hexadecapole
residual interactions on
rotational bands
 is investigated in a single-$j$ shell model.
An exact shell-model
diagonalization of the quadrupole-plus-hexadecapole Hamiltonian
demonstrates
that the hexadecapole-hexadecapole interaction can sometimes produce
a  staggering
of energy levels in  the yrast sequence;
however, long and regular $\Delta I$=2 sequences are not obtained.
The shell-model results are discussed in terms of the intrinsic
deformations extracted by means of the self-consistent Hartree-Fock
method. The angular momentum dependence of
intrinsic  quadrupole and hexadecapole
moments $Q_{2\mu}$ and $Q_{4\mu}$ is investigated.
\end{abstract}


\narrowtext

\section{Introduction}\label{introduction}

Recently, there has been  considerable interest in the
behavior of hexadecapole deformations at high spins,
motivated by the observation
of unexpected regular variations  in
the spectra of superdeformed (SD) bands. Namely, in several
SD  bands,
$^{149}$Gd\cite{[Fli93]}, $^{153}$Dy\cite{[Ced95]},
$^{194}$Hg\cite{[Ced94]},
$^{192}$Tl\cite{[Fis96]},
$^{148}$Gd\cite{[Ang96]},
 and $^{131,132}$Ce\cite{[Sem96]},
a $\Delta I$=2 staggering of the dynamical moment of inertia,
${\cal J}^{(2)}$, has been observed.
The effect is very weak and in some cases rather uncertain
\cite{noC4}.
It  manifests itself
in a systematic shift of 
 every other state in a rotational band. That is,
the sequence of states with spins $I$, $I$+4, $I$+8, etc.,
is shifted down with respect to the sequence $I$+2, $I$+6, $I$+10, etc.
In other words, the transition energies  of those bands
can be parametrized as:
\begin{equation}\label{C4}
E_\gamma(I)=\tilde{E}_\gamma(I)+\epsilon(I)(-1)^{\frac{I}{2}},
\end{equation}
where $\tilde{E}_\gamma(I)$, the average (reference)
$\gamma$-ray energy, and
the perturbation $\epsilon(I)$ are smoothly varying functions of $I$.
According to the data,
the amplitude of the staggering, $|\epsilon(I)|$,
 is very small -  of the order of 100 eV.

Since these oscillations separate a band into two families
in which spins differ
by  {\em four} units of angular momentum,
it seems natural to attribute their origin
to a coupling between the
rotational motion and hexadecapole
vibrations\cite{[Fli93],[Pek83]}.
This scenario requires the presence  of
a small component with a four-fold rotational symmetry, C$_4$,
in the mean field
of the nucleus.

The effect of such a term
has been investigated in phenomenological models
assuming  the C$_4$ symmetry axis coinciding either with the symmetry axis
($z$-axis) of the
quadrupole tensor\cite{[Ham94],[Mag95],[Mac95]} or with the
 rotation axis
($x$-axis)\cite{[Fli93],[Pav95],Mag96}.
These
 theoretical  studies   have assumed the presence of a
non-axial hexadecapole
term in the Hamiltonian of the rotating  nucleus.
However, microscopic calculations
based on the shell correction method\cite{[Rag94],[Fra94],[Don96],[Luo95]}
or on the self-consistent Hartree-Fock (HF)
method\cite{[Mag95a],[Bou96]}
predicted  an extremely   weak   collectivity
associated with the non-axial $\lambda$=4 fields.

In Ref.\cite{[Bur95a]},
in a simple quadrupole-plus-hexadecapole model which  does not
impose  {\em a priori} any intrinsic deformations,
it has been  demonstrated
that the staggering in  $E_\gamma$  may occur in certain cases.
The main objective of the present  study, based
on the same model,   is two-fold.
First, we analyze  the fluctuations (staggering) in the yrast
line  and discuss this phenomenon  in terms of
quadrupole and hexadecapole interactions. Second,
we analyze the angular momentum dependence of intrinsic
hexadecapole moments, $Q_{4\mu}$,
especially  those with $\mu$=2 and 4.
To the best of our knowledge, no systematic study of this
effect exists in literature. Also,
experimentally, very little  is known about
the  non-axial
hexadecapole deformations, $Q_{42}$ and $Q_{44}$,
and about their $I$-dependence.
Some limited evidence for these exotic deformations
exists from the
alpha scattering studies\cite{[Gov86],[Gov87]}
around $^{168}$Er.

The model to test the subtle phenomena discussed above  should
be capable of taking into account the interplay between rotation
and shape dynamics of a many-body system.
In our study, we
investigate a model which can be
solved exactly, namely, the model of a
single-$j$ shell
filled with  an even number of identical nucleons
interacting via the multipole forces.
This model,
in the  quadrupole-quadrupole  variant\cite{[Bel59]},
was widely exploited in the past to study nuclear collective effects
associated with quadrupole degrees of freedom.
In spite of its simplicity and a rather limited configuration space,
this model is able to
 describe a variety of collective phenomena including
 collective rotation. For instance,
even for as few as four  particles in a single-$j$ shell,
Mulhall and Sips\cite{[Mul64]}
found rotational structures present in the spectra.
Collective spectra for various particle numbers and
different values of  $j$
were  analyzed by Friedman and Kelson\cite{[Fri70]}.
Baranger and Kumar\cite{[Bar65],[Bar68]} studied
quadrupole deformability
and the influence of pairing, and
Arima\cite{[Ari68]} discussed excited rotational structures.
Recently, within the large-$j$ shell model,
the quadrupole collectivity was
studied by Burzy\'nski and Dobaczewski\cite{[Bur95]}.

The paper is organized as follows.
The single-$j$ shell model is briefly
 described in Sec.~\ref{models}.
Section \ref{exact}  contains the shell-model
results. In particular,  high-spin fluctuations in the yrast line
are discussed in terms of the interplay between the quadrupole and
hexadecapole components
 of the residual
 interaction. The corresponding intrinsic deformations
are studied in
Sec.~\ref{meanfield} within the self-consistent Hartree-Fock (HF) approach.
Finally, conclusions are contained in Sec.~\ref{conclusions}.

\section{The Model}\label{models}

The model studied
 in this paper is an extension of
a single-$j$ shell  model of Refs.
\cite{[Bel59],[Mul64]}. Its Hilbert space is that of
a  (large) single-$j$ shell
occupied by identical $N$ nucleons ($N$-even).
The model
 Hamiltonian is rotationally invariant and takes the form:
\begin{equation} \label{Ham}
    \hat{H}  =  \hat{H}_{\rm s.p.} + \hat{V}_{QQ},
\end{equation}
where
$\hat{H}_{\rm s.p.}$ is
the single-particle Hamiltonian, while
\begin{equation} \label{Ham1}
 \hat{V}_{QQ} =  -\bar{\chi}_2 \hat{Q}_2 \cdot \hat{Q}_2
             - \bar{\chi}_4 \hat{Q}_4 \cdot \hat{Q}_4
\end{equation}
 represents
the quadrupole-quadrupole and hexadecapole-hexadecapole
residual interaction.
For the single-$j$ shell, $\hat{H}_{\rm s.p.}$
reduces to a constant assumed to be zero in the
following.
The multipole operators appearing in Eq.~(\ref{Ham1})
 can be expressed as
\begin{equation} \label{Qop}
  \hat{Q}_{\lambda\mu} = \sum_{mm'}(jmjm'|\lambda\mu) a_m^+ \tilde a_{m'},
\end{equation}
where
$\tilde a_m \equiv \hat{T} a_m \hat{T}^{-1}$
($\hat{T}$ is the time reversal operator)
 and
the dot symbols in  Eq.~(\ref{Ham1}) denote the scalar product
\begin{equation} \label{Qdot}
  \hat{Q}_\lambda\cdot\hat{Q}_\lambda=
  \sum_{\mu}(-)^{\mu} \hat{Q}_{\lambda\mu}\hat{Q}_{\lambda,-\mu}.
\end{equation}

The Hamiltonian (\ref{Ham},\ref{Ham1}) is fully described by
the two  constants  $\bar{\chi}_2$ and $\bar{\chi}_4 $ representing
the  strengths of the quadrupole and hexadecapole interactions, respectively.
The dimension of the corresponding  Hilbert space is finite.
Consequently, the exact solutions of this quantum many-body problem
can be obtained by means of a  straightforward
diagonalization \cite{[Bur95]}. This has been done in three steps, as
described in the following.
First,
the $M$-scheme basis was constructed. It corresponds to
$N$-particle wave functions $|NMn\rangle$, with
 $M$ being  the projection of the total angular momentum $I$,
and $n$ distinguishing  between states with the same value of $M$.
Since the model Hamiltonian is rotationally invariant, its energies
do not depend on $M$. Hence
$M$ was assumed to be equal to zero.
Second, the $I$-scheme basis
 $|NIk\rangle$ ($M$=0) was  obtained by
numerically diagonalizing
the operator $\hat{I}^{2}$ ($\hat{I}$ is the total
angular momentum of the system) in the $M$-scheme basis.
Here, the
additional quantum number $k$ distinguishes between
states with the same angular momentum $I$.
Finally,  the Hamiltonian has been diagonalized in the $I$-scheme
basis.

{}For the half-filled shell
there appears an additional quantum number related to the
particle-hole symmetry
associated with the unitary symmetry operator
\begin{equation} \label{sym}
 \hat{\cal U}=\prod_{m=-j}^{j} (a^{+}_{m} + a_{m}).
\end{equation}
[An arbitrary but fixed order of factors in
Eq.~(\ref{sym}) is assumed.]

Since  $\hat{\cal U}^2$=$(-1)^{j+1/2}$,
depending on the value of $j$,  $\hat{\cal U}$
 is hermitian or antihermitian.
By employing the identities:
\begin{equation} \label{symq}
 \hat{\cal U}a^{+}_{m} =  -a_{m} \hat{\cal U},
 ~ \hat{\cal U}a_{m} =  -a^+_{m} \hat{\cal U},
\end{equation}
it is easy to show that the particle-number operator
$\hat N$ transforms
under $\hat{\cal U}$ as
\begin{equation} \label{symp}
 \hat{\cal U}^+\hat{N}\hat{\cal U} =
2j+1 - \hat{N},
\end{equation}
i.e., $\hat{\cal U}$ transforms a
state with $N$ particles into a state
with $2j$+1$-$$N$ particles.
Moreover, because of the relation
\begin{equation} \label{Ptran}
  \hat{\cal U}^+\hat{Q}_{\lambda\mu}\hat{\cal U}
= -(-1)^\mu\hat{Q}_{\lambda,-\mu}~~(\lambda>0),
\end{equation}
$\hat{\cal U}$ commutes with  Hamiltonian (\ref{Ham1}).
Consequently,
 the spectra of $N$- and ($2j$+1$-$$N$)-particle systems
are exactly degenerate, while for the half-filled shell,
$N$=$j$+1/2, the
eigenstates of Hamiltonian (\ref{Ham})
acquire a new quantum number $u$=$\pm i^N$,
an eigenvalue of the $\hat{\cal U}$ operator.

Properties of  Hamiltonian (\ref{Ham})
depend only on the ratio $\chi_4$=$\bar{\chi}_4/\bar{\chi}_2$. The
absolute value of $\bar{\chi}_2$ determines the energy scale of
the model. Consequently, in the following, we put $\bar{\chi}_2$=1 MeV.
The ``realistic" value
of $\chi_4$ can be estimated by means of the self-consistency relation
of Refs.\cite{[Row67],[Boh75],[Sak89]}.
In the harmonic oscillator approximation for the
average potential and assuming the nuclear radius to be
$R$=1.2$A^{1/3}$\,fm, the
self-consistent ratio $\chi_4$ is equal to 165/176, i.e.,
it is very close to unity. Therefore,
in our study the value
of $\chi_4$ was varied from 0 (no hexadecapole interaction) to 1
(self-consistent limit).

The intrinsic deformations were obtained by means of the
standard cranking HF procedure.
The self-consistent one-body density
 in the rotating frame, $\hat\rho^\omega$,
was found by solving the cranking HF equations
\begin{equation}\label{crank}
\left[\hat{h}^\omega(\hat\rho^\omega),\hat\rho^\omega \right]=0,
\end{equation}
where $\hat{h}^\omega$ is the self-consistent Routhian,
$\hat{h}^\omega$=$\hat{h} - \omega \hat{I}_x$,
$\hat{h}$=Tr$(\hat{V}_{QQ}\hat\rho^\omega)$
is the HF Hamiltonian
(including the exchange term),
$\hat{I}_x$ is the component of the total angular momentum
along the $x$ axis, and
$\omega$ is the rotational frequency determined from the
angular momentum equation:
Tr$(\hat\rho^\omega \hat{I}_x)$=$I$.

The maximum angular momentum carried out by a  system of
 $N$ particles in a single-$j$ shell is
\begin{equation}\label{Imax}
I_{\text{max}} = \frac{N(2j+1-N)}{2},
\end{equation}
e.g., it is largest for a half-filled shell
[$I_{\text{max}}$=$(2j+1)^2/8$]. For instance, for $j$=19/2
and $N$=10, $I_{\text{max}}$=50.

Having found the self-consistent density matrix, the intrinsic
multipole moments,
\begin{equation}\label{defs}
Q_{\lambda \mu} = {\rm
Tr} (\hat \rho^ \omega \hat{Q}_{\lambda\mu}), ~~~(\lambda=2, \, 4)
\end{equation}
were calculated as functions of $I$. [The multipole operators
$\hat{Q}_{ \lambda \mu}$ appearing in Eq.~(\ref{defs}) are those of
Eq.~(\ref{Qop}).] Since it is assumed that the self-consistent density
matrix has three symmetry planes (i.e., $D_{\rm 2h}$ is a self-consistent
symmetry), the odd-$\mu$ components of $Q_2$ and $Q_4$ vanish,
and $Q_{\lambda \mu}$=$Q_{\lambda- \mu}$. This is consistent with the standard
definition of the intrinsic system defined in terms of the principal axis
of quadrupole tensor.

It is convenient to relate the moments $Q_{\lambda\mu}$
($\lambda$=2, 4) to simpler deformation parameters that guarantee
the unique mapping of the $Q$-surface on the parameter values.
The
 two quadrupole moments, $Q_{20}$ and $Q_{22}$,
are usually written in terms of the ``polar"
coordinates $\beta_2$ and $\gamma_2$\cite{[Boh52],[Hil53]}:
\begin{equation}\label{quadr}
Q_{20}=\beta_2\cos\gamma_2,
 ~~~Q_{22} =
 \frac{1}{\sqrt{2}}\beta_2\sin\gamma_2.
\end{equation}
Since in our case $\hat\rho^\omega$ has three symmetry planes,
the hexadecapole tensor can be parametrized
by means of the three
``spherical" coordinates, $\beta_4$, $\gamma_4$, and $\delta_4$
\cite{[Roh81]}:
\begin{eqnarray}\label{q4def}
Q_{40} & = & \beta_4\left(\sqrt{{7\over{12}}}\cos\delta_4
+\sqrt{{5\over{12}}}\sin\delta_4\cos\gamma_4\right), \nonumber \\
Q_{42} & =
& {1\over{\sqrt{2}}}\beta_4\sin\delta_4\sin\gamma_4, \\
Q_{44} & = & \beta_4\left(\sqrt{{5\over{24}}}\cos\delta_4
-\sqrt{{7\over{24}}}\sin\delta_4\cos\gamma_4\right), \nonumber
\end{eqnarray}
where 0$\le$$\delta_4$$\le$$\pi$ and
$-\pi$$\le$$\gamma_4$$\le$$\pi$.
Since in definition (\ref{Qop}) the radial dependence of $\hat{Q}$
has been ignored (the radial matrix
element  is the same for all magnetic
substates, i.e., it is a scaling factor),
the overall scale of the deformation variables
$\beta_2$ and $\beta_4$ is given by the magnitude of the quadrupole
operators (\ref{defs}), and is not related to the scale of the
usual Bohr's shape parameters\cite{[Boh52]}.

The  hexadecapole shape that is axial with respect to the $z$-axis
corresponds to $\gamma_4$=$\gamma_{4}^{(z)}$=0 
and $\delta_4$=$\delta_4^{(z)}$,
where $\cos\delta_4^{(z)}$=$\sqrt{7/12}$
($\delta_4^{(z)}$$\simeq$0.7017 or 40$^\circ$20'). Analogously,
the hexadecapole shape which is axial
with respect to the $x$-axis
corresponds to $\gamma_4$=$\gamma_{4}^{(x)}$=2$\pi$/3
and $\delta_4^{(x)}$=$\delta_4^{(z)}$.

\section{Shell Model Results}\label{exact}

To analyze the staggering effect in collective bands, several
criteria  have been introduced in the literature.  One
possibility was suggested in  Ref.\cite{[Fli93]} where the
staggering was discussed in terms of
\begin{equation}\label{flibotte}
\Delta^3E_\gamma(I) \equiv
\frac{1}{4}\left[
E_\gamma(I-3)-3E_\gamma(I-1)+3E_\gamma(I+1)-E_\gamma(I+3)
\right],
\end{equation}
while  another staggering filter,
\begin{equation}\label{cederwall}
\Delta^4E_\gamma(I) \equiv
\frac{3}{8}\left[
E_\gamma(I)-\frac{1}{6}\left(4E_\gamma(I-2)+4E_\gamma(I+2)-
E_\gamma(I-4)-E_\gamma(I+4)
\right)\right],
\end{equation}
was introduced in Ref.\cite{[Ced94]}.  The usefulness of
quantities (\ref{flibotte}) and (\ref{cederwall}) for
visualizing irregularities in rotational bands has recently been
discussed in Ref.\cite{[Jin95]} where it was concluded that, in
some cases, they can dramatically overemphasize perturbations in
rotational levels (which occur, for instance, due to accidental
degeneracies).  In this context, we would like  to point out that
 one can construct a set of quantities $\Delta^kE_\gamma$,
related to various derivatives of $E_\gamma$ with respect to
$I$. Indeed, for a function $f(x)$ defined on a discrete
equidistant grid,  $f_i$=$f(x_i)$ ($x_{i+1}-x_i=h)$, its
derivatives can be approximated by the finite difference
formulas:
\begin{mathletters}
\label{deriv}\begin{eqnarray}
 \left(\frac{df}{dx}\right)_{x=(x_i+x_{i+1})/2}  &\approx&
\frac{1}{h}\left(f_{i+1}-f_i\right) \label{deriv-a} \\[2ex]
  \left(\frac{d^2f}{dx^2}\right)_{x=x_{i}}~~~~~~~~~~ &\approx&
\frac{1}{h^2}\left(f_{i-1}-2f_{i}+f_{i+1}\right) \\[2ex]
  \left(\frac{d^3f}{dx^3}\right)_{x=(x_i+x_{i+1})/2}  &\approx&
\frac{1}{h^3}\left(3f_{i}-3f_{i+1}-f_{i-1}+f_{i+2}\right) \\[2ex]
 \left(\frac{df^4}{dx^4}\right)_{x=x_i}~~~~~~~~~~  &\approx&
\frac{1}{h^4}\left(6f_i
-4f_{i+1}-4f_{i-1}+f_{i-2}+f_{i+2}\right).
\end{eqnarray}
\end{mathletters}
It is easy to see that  filters (\ref{flibotte}) and
(\ref{cederwall}) can be expressed as
\begin{equation}
\Delta^3 E_\gamma \approx -2 \left(\frac{d^3E_\gamma}{dI^3} \right)
 ~~~{\rm and}~~~~\Delta^4 E \approx  \left(\frac{d^4E_\gamma}{dI^4} \right).
\end{equation}
That is, $\Delta^3 E_\gamma$ and $\Delta^4 E_\gamma$ are
related, respectively, to the fourth and fifth derivative of the
total energy, $E$,  with respect to $I$.
By going to higher derivatives in defining the staggering
filter, one effectively spreads out
the perturbation due to configuration mixing
at  $I$=$I_0$ over many states around $I$=$I_0$.
Consequently, in order to avoid this artificial effect,
in the present study we employ the simplest filter,
\begin{equation}\label{baktash}
\Delta^2E_\gamma(I) \equiv
\frac{1}{4}\left[
E_\gamma(I-2)-2E_\gamma(I)+E_\gamma(I+2)\right],
\end{equation}
as a measure of fluctuations \cite{[Bak95]}.
According to Eq.~(\ref{deriv-a}),  $\Delta^2 E_\gamma$
is the second derivative of $E_\gamma$ with respect to $I$, or
\begin{equation}
\Delta^2 E_\gamma \approx 2 \frac{d}{dI} \frac{1}{{\cal J}^{(2)}}
=-2\frac{1}{({\cal J}^{(2)})^2}
\frac{d{\cal J}^{(2)}}{dI},
\end{equation}
where ${\cal J}^{(2)}$ is the second moment of inertia.
It is worth noting that for a ``C$_4$" spectrum, [Eq.~(\ref{C4})] with
$\tilde{E}_\gamma$ linear in $I$ (perfect rotor) and
$\epsilon(I)$=$\epsilon$=const., all filters $\Delta^k E_\gamma$
($k$=2, 3, 4) give results proportional to $\epsilon$.

We now proceed by applying the staggering filter (\ref{baktash})
to the results of the exact shell model diagonalization of
the single-$j$ shell
Hamiltonian (\ref{Ham}).
Figure~\ref{spectra8}
displays the even-$I$ yrast
lines for $N$=8 and 10
particles
 moving in the $j$=19/2 shell, and interacting with
the quadrupole and hexadecapole interactions
with $\chi_4$=0, 0.5, and 1. For the
half-filled $j$=19/2 shell ($N$=10)
the particle-hole
symmetry $\hat{\cal U}$ is preserved and
the open full
dots indicate states having
 $u$=1 and $u$=$-$1, respectively. The corresponding values of
 $\Delta^2E_\gamma$ are presented in
Fig.~\ref{stagger8}.

In both cases, $N$=8 and 10, the yrast line at
$\chi_4$=0 and 0.5 can be understood in terms
of  smooth collective  bands
connected by large stretched ($\Delta I$=2) transition matrix
elements of the
quadrupole operator, Eq.~(\ref{Qop}).
 In the limit of the strong hexadecapole
interaction, $\chi_4$=1, the yrast line becomes highly perturbed
and rather irregular; the quadrupole collectivity is lost, i.e.,
the yrast states are no longer connected by large quadrupole
matrix elements.

{}For the half-filled shell,
the yrast line obtained in
  the $\chi_4$=0 and $\chi_4$=0.5 variants
consists of states with alternating values of $u$.
Thanks to relation
(\ref{Ptran}), multipole operators only connect
states with different $u$'s. Consequently,
for small  and intermediate  values of $\chi_4$,
the yrast line consists of  long {\em quadrupole
band sequences}, i.e., states
connected by the  $Q_2$ operator, while the
stretched ($\Delta{I}$=4) hexadecapole transitions
are strictly forbidden. The situation changes dramatically
for large relative strengths $\chi_4$. Here, due to strong hexadecapole
correlations favoring the $\Delta I$=4 coupling, the alternating-$u$
pattern breaks down and  the neighboring states often
 have the same
values of $u$, i.e., they are not connected by a multipole operator.
Hence, the yrast line cannot  be considered as a rotational band.
For the $N$=8 case, $u$ is not conserved, and the
$u\rightarrow -u$ selection rule for the transition operator
does not hold.

By comparing Fig.~\ref{spectra8} with
Fig.~\ref{stagger8}, one can see that the
staggering filter (\ref{baktash}) singles out all kinds of
irregularities occurring along the yrast line.  For example, the
characteristic fluctuation   at $I$=16
 and $\chi_4$=0 (for both $N$=8 and 10)
 results from a small kink in the yrast line
caused by  a band crossing. Due to large band interaction, this
irregularity is
barely visible in Fig.~\ref{spectra8}.

According to Fig.~\ref{stagger8},
values of $\Delta^2E_\gamma$ increase with  $\chi_4$.
 For $\chi_4$=1 they are an order of magnitude larger
than for $\chi_4$=0.5. Relatively long sequences of regularly
staggered points are visible for $\chi_4$=1 at large values of
the angular momentum. One has to note, however,
that the quantity  $\Delta^2E_\gamma$ represents the staggering
along the yrast line, which in the $\chi_4$=1 case does not correspond to a
sequence of rotational bands.
 For $\chi_4$=0.5, the magnitude of staggering
is  smaller and the staggering sequences are much shorter.

As discussed above, the yrast
lines calculated for $N$=10, and $\chi_4$=0 and $\chi_4$=0.5 do
consist of two  alternating $\Delta I$=4
sequences with the same values of $u$.
 However, this fact is not
reflected in the corresponding staggering pattern. On the other
hand, the yrast band for $\chi_4$=1 does not have a form of an
alternating-$u$ structure,  but  the staggering pattern is
clearly seen. It is, therefore, obvious that the sole
existence of two
different representations of states with opposite values
of $(-1)^{I/2}$ is not a sufficient condition for the appearance
of the staggering pattern.  In any case, the hexadecapole
interactions considered in our study do not differentiate
between the two particle-hole symmetries which occur for the
half-filled shell.

{}From the shell-model results alone,  one is not able to
understand  the origin
of the longer or shorter staggering patterns obtained in the
calculations. In order to analyze the underlying physics,
 we proceed in the next section by investigating
the mean-field properties of the single-$j$ shell model.

\section{Mean-Field Approximation}\label{meanfield}

The cranking HF equation (\ref{crank}) was solved for the
Hamiltonian (\ref{Ham},\ref{Ham1}) and  for the same values of $j$, $N$, and
$\chi_4$ as used in the shell-model study.
At the largest values of $\omega$, the HF solutions correspond
to the fully aligned states, i.e., they are given by the Slater
determinants of fermions occupying single-particle states with the 
largest available projections of the angular momentum on the
rotation axis. This is illustrated in Fig.~\ref{routhians} which
displays
 the single-particle routhians (the eigenvalues
of Routhian,
$\hat{h}^\omega|\nu\rangle$=$e^\omega_\nu|\nu\rangle$) and the
corresponding single-particle angular-momentum
alignments
$i_\nu$.  We see that for the
large rotational frequencies, the routhians can be represented by
straight lines
with slopes given by constant alignments.  (Recall that in
the HF theory the relation $i_\nu$=$-de^\omega_\nu/d\omega$ holds
only approximately because of the $\omega$-dependence of the
average HF field.)

The total alignments $\langle\hat{I_x}\rangle$=$\sum_\nu i_\nu$
(the average values of $\hat{I}_x$ in the many body HF states)
are shown in Fig.~\ref{iom8} for the same
parameters
as  used  in
Figs.~\ref{spectra8} and
\ref{stagger8}. For $N$=8 and
$\chi_4$=0 or 0.5, the limit of a  full alignment
 $I_{\text{max}}$=48
is reached.
A conspicuous feature of the cranking HF results presented in
Fig.~\ref{iom8}  is the appearance of gaps in
the calculated yrast line. This can be attributed to the
old problem in the description of band crossing in terms of the cranking
model \cite{[class]}. Namely, in the region of band crossing
 the  lowest
self-consistent  solution
jumps as a function of $\omega$ from one continuous family of
states to another.

In  cases shown in Fig.~\ref{iom8},
a band crossing occurs
for  $\chi_4$=0 and 0.5 (and for both values of $N$). On the other hand,
for $\chi_4$=1 there appear many consecutive band crossings
related to configuration changes, and
the yrast line is  composed of short pieces of continuous HF
solutions.  This effect reflects the fact discussed in
Sec.~\ref{exact}
 that for large values of $\chi_4$ the yrast states
of the exact solution cannot be grouped in bands
 connected by strong
quadrupole transition matrix elements.  It also explains the
strong irregularities seen in the exact yrast lines
calculated  with large
hexadecapole interactions. When these irregularities are
analyzed with
 a staggering filter, one may obtain sequences of
staggered points; however, in the single-$j$ shell model studied
here this does not seem to be connected with any
collective effect caused by the hexadecapole interaction.
On the other hand, even an isolated band crossing, as seen in the HF
results for $\chi_4$=0 and 0.5, is reflected by a kink in the
exact shell-model yrast line, i.e.,
short deviations of $\Delta^2 E_\gamma$ from zero
neatly exposed by
the staggering filter, cf.\ Fig.~\ref{stagger8}.

The HF energies  are compared in
Fig.~\ref{spectra8} with  the results of the
exact shell-model diagonalization. One can see that for
$\chi_4$=0 the mean field approximation reproduces the exact
results   well \cite{[Bur95]}.  Also for $\chi_4$=0.5 the
approximation is fair, although the exact results visibly
deviate at low angular momenta from a parabolic, rotor-like
behavior given by the HF solutions. In particular,
for  both values of $\chi_4$,
band crossings  appear in the HF curves very close to
the  kinks seen
in the exact yrast spectra.

On the other hand,
for $\chi_4$=1 the HF approximation fails to
reproduce the exact states at low angular momenta. Interestingly,
both shell-model and HF  predict the existence of yrast traps
at $I$$\approx$16; the multiple band
crossing phenomenon is clearly correlated with
the staggered
shell-model energies.

At very high angular momenta, when the full alignment limit is
reached, the HF results approach the exact solution. Indeed,
for $I$=$I_{\text{max}}$, the shell-model wave function is
represented by a single,
 fully aligned Slater determinant, a HF state.

Based on the HF solutions, we may now discuss the nature of
crossing  bands. To this end, we show
in Figs.~\ref{alpha2_8} and
\ref{alpha4_8}
 the average values of
quadrupole and hexadecapole moments [Eq.~(\ref{defs})],
 respectively.  One can see that
the deformation patterns for $\chi_4$=0 and 0.5 are rather
similar, and these patterns undergo qualitative changes when $\chi_4$
increases to 1.
Since our aim is to discuss collective rotation, in the following
we concentrate on the case of
low hexadecapole strengths.

 Figure~\ref{triangle} shows  the values of quadrupole
 deformation  $\beta_2$ and $\gamma_2$
[Eq.~(\ref{quadr})]
 for the HF yrast lines at  $\chi_4$=0.5. It is to be noted that
in our paper we consequently employ the Bohr-Hill-Wheeler
 convention of
$\gamma_2$ deformation, not the standard Lund convention.
Consequently, the collective prolate rotation takes place
at  $\gamma_2$=0$^\circ$ (the $z$-axis is the symmetry axis)
and  $\gamma_2$=--120$^\circ$ (the $y$-axis is the symmetry axis),
while the  collective oblate rotation corresponds to
 $\gamma_2$=60$^\circ$ (the $y$-axis is the symmetry axis)
or  $\gamma_2$=$\pm$180$^\circ$ (the $z$-axis is the symmetry axis).
The non-collective rotation around the $x$-axis corresponds to
 $\gamma_2$=--60$^\circ$ (oblate shape) and  $\gamma_2$=120$^\circ$
(prolate shape). The regions ($-60^\circ\leq\gamma_2\leq 120^\circ$)
and ($\gamma_2\leq-60^\circ$ and $\gamma_2\geq 120^\circ$) are physically
equivalent.

As seen in  Fig.~\ref{triangle},   the results for  $N$=8 and 10
are very similar. Namely, the quadrupole deformation
$\beta_2$ decreases with increasing angular momentum.
This
decrease is almost continuous across the crossing point at about
$I$=16. On the other hand, at the crossing point there is a
sudden change of the $\gamma_2$ values from about $-180^\circ$
(collective oblate rotation
to about $-150^\circ$ (maximally triaxial shape  with
the intermediate axis being the $x$ axis).

In the fully aligned state, $I$=$I_{\text{max}}$, the equilibrium shape
of a system with
 $N$$<$$j$+1/2 corresponds  to $\gamma_2$=$\gamma_{2}^{(x)}$=$-60^\circ$.
This is so because the eigenstates of $I_x$ cannot have nonzero
average values of $\hat{Q}_{2,\mu_x=2}$,
 i.e., the $x$ axis must be the symmetry axis, and,
moreover, for
 $N$$<$$j$+1/2
the shape must be oblate.
In this limit, values of quadrupole
moments along the $z$ axis are proportional to the $Q_{20x}$ values
calculated along the $x$ axis. Namely:
$\displaystyle{Q_{20}=-\frac{1}{2}Q_{20x}}$,
$\displaystyle{Q_{22}=\frac{1}{2}\sqrt{\frac{3}{2}}Q_{20x}}$,
and $\beta_{2}=|Q_{20x}|$.
This is seen in Fig.~\ref{triangle} for $N$=8 at the
high-angular-momentum end of the band.
On the other hand, for the half-filled
shell the aligned state has $\beta_2$=0, and this point is
approached along the
$\gamma_2$=$-150^\circ$ line.

Of course, this description in terms of the shape
characteristics pertains only to the average values of the
quadrupole moments, $Q_{20}$ and $Q_{22}$, which have
well-known values for ellipsoidal shapes.  In the following, we
refer to the bands below and above the crossing as the
collective oblate
and triaxial bands, respectively, and the band crossing can
be interpreted as a sudden transition caused
by a change in the shape of the system.

We may now discuss the evolution of the hexadecapole shape of
the system within the oblate and triaxial bands.  The oblate
bands do not exhibit any tangible non-axial hexadecapole
deformations, i.e., they have
$Q_{42}$$\approx$$Q_{44}$$\approx$0.  On the other hand, in the
triaxial bands all three components of the hexadecapole tensor are
nonzero. In both bands the hexadecapole deformations
 are fairly similar for $\chi_4$=0 and 0.5, i.e., they weakly
depend on the fact whether the hexadecapole interaction has or
has not been
taken into account. This suggests a simple interpretation of
the hexadecapole moments as being induced by the quadrupole
deformations. Indeed, ellipsoidal shapes
 have nonvanishing  hexadecapole moments which can be expressed through
quadrupole moments or
deformations. This
leads to rather stringent conditions for the hexadecapole
moments \cite{[Roh74],[Naz81],[Roh96]}.

The assumption of the strong coupling of $\lambda$=4
deformations to the intrinsic frame defined through quadrupole
deformations requires that the hexadecapole tensor depends on
 products of quadrupole tensors. This leaves a freedom of
many possible parametrizations\cite{[Roh96]}
defined through three scalar functions $h_2$, $h_3$, and $h_4$,
depending on $\beta_{2}$ and $\cos3\gamma_2$. In the case that
only $h_3$ is different from zero, one obtains:
\begin{mathletters}\label{roh1}
\begin{eqnarray}
\tilde\beta_4 & = & |h_3|\sqrt{
         \left(\frac{5}{12}+\frac{7}{12}\cos^{2}3\gamma_{2}
\right)}, \label{roh-a}  \\
\cos\tilde\delta_{4} & = & \pm\frac{\cos3\gamma_{2}}
           {\sqrt{\frac{5}{7}+\cos^{2}3\gamma_{2}}},
 \label{roh-b}  \\
\cos\tilde\gamma_{4}  =  \pm \cos\gamma_{2},
 &~~&\sin\tilde\gamma_{4}  =  \pm \sin\gamma_{2}  \label{roh-c}.
\end{eqnarray}
\end{mathletters}
[There are two possible sign choices in Eqs.~(\ref{roh-b}) and (\ref{roh-c}):
either $\tilde{\gamma}_{4}=\gamma_{2}$ or
$\tilde{\gamma}_{4}=\gamma_{2}\pm\pi$.]

Expressions (\ref{roh1}) define shapes for which the hexadecapole
moments follow the quadrupole moments and do not constitute any
independent dynamical deformations.
(For the general parametrization for the  hexadecapole tensor, see
Ref.~\cite{[Roh96]}.)
According to  Eq.~(\ref{roh-a}) the value of  $|h_3|$ can
 always be chosen in such
a way that the strong-coupling and self-consistent values of $\beta_4$
are equal, i.e., $\beta_4$=$\tilde\beta_4$. On the other hand,
the strong-coupling
values of angles $\tilde\delta_4$ and  $\tilde\gamma_4$ depend
solely on $\gamma_2$.

  It turns out that the HF
results obtained for low and intermediate values of $\chi_4$
can be understood in terms of the
strong-coupling limit given by Eq.~(\ref{roh1}).
Figure~\ref{def4_8} shows the
calculated HF  hexadecapole deformations $\beta_4$,
 $\delta_4$, and $\gamma_4$
 together with the strong-coupling values
of  $\tilde\delta_4$ and  $\tilde\gamma_4$ given by
Eqs.~(\ref{roh-b}) and (\ref{roh-c}), respectively.
 On can see that
the  assumption  of a strong coupling
accounts very well for the complicated
$I$-dependence of $\delta_4$ and $\gamma_4$
 presented in Fig.~\ref{alpha4_8}.
Moreover one should notice that for the case of $N$=8 the values
of $\delta_{4}$ and $\gamma_{4}$ are approaching the limit
of the fully aligned state along $x$-axis where only the value
of $\hat{Q}_{40x}$ remains nonzero.

It is interesting to note that in the $N$=10 variant and $I$=50
(termination point), hexadecapole deformation $\beta_4$
vanishes for $\chi_4$=0.5. Indeed, in the limit of angular momentum
alignment, the expectation value $\langle jm_x|Q_{\lambda\mu}|jm_x\rangle$
is proportional to
$d^\lambda_{\mu0}(\frac{\pi}{2})\langle jm_x\lambda 0|jm_x\rangle$, i.e.,
it depends only on $m_x^2$ for even values of $\lambda$
[$\langle jm_x\lambda 0|jm_x\rangle \approx
P_\lambda\left(\frac{m_x}{j}\right)$ and for large values of $j$].
Hence the HF expectation
values $Q_{4\mu}$ vanish in the fully aligned state of the half-filled shell
($\int_1^0 P_\lambda(x)dx =0$ for even values of $\lambda$).

In the case of $N$=8 particles,
the average values of $\hat{Q}_{4\mu}$
can be expressed by the axial hexadecapole moment
calculated along the $x$ axis.
Namely, the following relations hold:
$\displaystyle{Q_{40}=\frac{3}{8}Q_{40x}}$,
$\displaystyle{Q_{42}=-\frac{1}{8}\sqrt{10}Q_{40x}}$,
$\displaystyle{Q_{44}=\frac{1}{8}\sqrt{\frac{35}{2}}Q_{40x}}$,
and $\beta_{4}=|Q_{40x}|$. Consequently, for $I$=48 the parameters
$\delta_4$ and $\gamma_4$ are approaching the values of $\delta_{4}^{(x)}$
and $\gamma_{4}^{(x)}$, respectively.

We conclude this section by presenting in Fig.~\ref{b4} the
hexadecapole deformation $\beta_4$ for $N$=8 and 10 as a
function of the coupling constant $\chi_4$ at $I$=40, i.e., for
the triaxial bands. On can see that for both
particle numbers, $\beta_4$  steadily
increases with $\chi_4$ in the whole range of $\chi_4$. That is,
the  hexadecapole phase transition
that would manifest itself in  a rapid local increase
of $\beta_4$ with $\chi_4$, is not present in the
single-$j$ shell model.

\section{Conclusions}\label{conclusions}

In the present paper we studied the single-$j$ shell model
describing identical
particles interacting via the quadrupole and hexadecapole
residual interaction. The model was solved exactly,  and the resulting yrast
structures  obtained for different  strengths of the hexadecapole
interaction were analyzed. The model was also solved
 within the Hartree-Fock
approximation, where the results could be interpreted in terms
of the standard shape variables.
Results of the calculations were presented for the  $j$=19/2 shell
and particle numbers $N$=8 and 10.  Calculations
 have also been performed
 for other values of  $j$ and $N$,
 but the main conclusions can be drawn from the
restricted set of results presented in the paper.

The purpose of our study was two-fold. First, we analyzed the
conjecture that the hexadecapole $\lambda$=4 degree of freedom
might be responsible for the $\Delta{I}$=2 staggering effect in
rotational bands. Second, we investigated the evolution of the
hexadecapole shapes with  angular momentum.

Our study shows that
 by including the
hexadecapole interaction in the single-$j$ model,
one can sometimes obtain a staggering of yrast energies.
 However, a
long and regular sequence of $\Delta{I}$=2
staggered energies cannot be obtained. This negative
result suggests that the experimental data 
probably cannot be explained in terms of the
 coupling between  rotation and hexadecapole
vibrations.

{}For a relatively weak strength of the hexadecapole interaction,
one obtains regular collective quadrupole bands which do not
exhibit any staggering.
 One also systematically obtains the effect of a
crossing between two different shape configurations:  an oblate-shape
collective
band at low angular momenta and a triaxial band at high angular
momenta. The irregularities which are present in the yrast
line
in the crossing region should not be confused with the
staggering phenomenon.
As discussed in Ref.~\cite{[Jin95]},
 such a misinterpretation may easily happen
especially when a multi-point staggering filter is used to
extract the staggering amplitude from the calculated energies.

{}For a relatively strong hexadecapole interaction, the yrast line
becomes rather irregular and the states are no longer
connected by strong quadrupole transition matrix elements. In
the mean-field picture, such a yrast structure can be interpreted as
composed of several different bands crossing one another.
In some cases discussed in this paper,
such  multiple crossings can give rise to
 staggering patterns.

In the studied model, one obtains  nonzero hexadecapole moments
even without including the hexadecapole interaction.  These
moments are rather weakly affected by
the  hexadecapole interaction.
The hexadecapole deformations calculated in the single-$j$ model
 simply follow the quadrupole
deformations, i.e., they can be discussed as resulting from a
strong coupling between the $\lambda$=2 and $\lambda$=4 modes.
Within the collective oblate bands, the hexadecapole moment is axial and
oriented along the symmetry axis of the quadrupole deformation.
For triaxial bands, the hexadecapole deformation also becomes 
triaxial, i.e., all three components of the hexadecapole moment
differ from zero. They can be nicely interpreted
within the strong coupling assumption, and depend in a simple
geometrical way on the quadrupole asymmetry angle $\gamma_2$.
It is worth noting that our self-consistent results strongly prefer
one particular parametrization of the hexadecapole tensor
of Ref.~\cite{[Roh96]}, namely the strong-coupling expression with
$h_2$=$h_4$=0.

\acknowledgements

This research was supported in part by the Polish Committee for
Scientific Research under Contracts Nos.~2P03B 098 09 and
2~P03B~034~08,  by a computational grant from the
Interdisciplinary Centre for Mathematical and Computational
Modeling (ICM) of  Warsaw University, and
by the U.S. Department of
Energy (DOE) through Contract No.
DE-FG02-96ER40963
 with
the University of
Tennessee.
Oak Ridge National
Laboratory is managed for the U.S. DOE by Lockheed
Martin Energy Research Corp. under Contract No.
DE-AC05-96OR22464.


\begin{figure}
\caption{\label{spectra8}
The exact yrast spectrum (circles) of the
quadrupole-plus-hexadecapole model for the $j$=19/2 shell filled
with $N$=8 (left) and 10 (right)
 particles, and  $\chi_4$=0 (top), 0.5 (middle), and 1
(bottom).  The yrast bands calculated in the HF method are
indicated by solid lines.
For $N$=10,
open and full circles correspond to
the values  $u$=1 and
$u$=$-$1, respectively, of the particle-hole symmetry quantum
number.
}
\end{figure}

\begin{figure}
\caption{\label{stagger8}
Staggering parameter $\Delta^2 E_\gamma$
[Eq.~(\protect\ref{baktash})] as a function of $I$
along  the
yrast lines obtained in the shell-model
results presented in  Fig.~\protect\ref{spectra8}.
}
\end{figure}

\begin{figure}
\caption{\label{routhians}
Top: Self-consistent single-particle routhians, $e^\omega_\nu$,
for $j$=19/2 and $N$=10 (yrast line) as functions of the rotational
frequency $\omega$.  Bottom:
the  corresponding single-particle alignments.
 Solid
lines: signature $r=+i$. Dashed lines: $r=-i$.}
\end{figure}

\begin{figure}
\caption{\label{iom8}
Angular momentum (the expectation value of $\hat{I}_{x}$ in the
cranking HF state) as a function of $\omega$ for
the $j$=19/2 shell filled with $N$=8
(left) and 10 (right) particles, and  $\chi_4$=0
(top), 0.5 (middle), and 1 (bottom).
}
\end{figure}

\begin{figure}
\caption{\label{alpha2_8}
Equilibrium quadrupole moments $Q_{20}$ (solid line)
and $Q_{22}$ (dot-dashed line)
for HF yrast solutions
as functions of angular
momentum for the $j$=19/2 shell with  $N$=8 (left)
and 10 (right) particles,  and  $\chi_4$=0
(top), 0.5 (middle), and 1 (bottom).
}
\end{figure}

\begin{figure}
\caption{\label{alpha4_8}
Same as in Fig.~\protect\ref{alpha2_8} except  for
equilibrium hexadecapole
moments  $Q_{40}$ (solid line)
and $Q_{42}$ (dot-dashed line), and $Q_{44}$
(dotted line).
}
\end{figure}

\begin{figure}
\caption{\label{triangle}
Equilibrium deformations $\beta_2$ and $\gamma_2$
[Eq.~(\protect\ref{quadr})] as functions
of  angular momentum for the HF yrast states   obtained in the
$j$=19/2 shell filled with $N$=8 (solid line) or $N$=10
(dot-dashed line) particles, and for $\chi_4$=0.5.
The limit of axial quadrupole shapes are indicated by dotted lines
($\gamma_{2}^{(x)}$: axial symmetry with respect to $x$-axis;
 $\gamma_{2}^{(z)}$: axial symmetry with respect to $z$-axis).
}
\end{figure}

\begin{figure}
\caption{\label{def4_8}
Equilibrium hexadecapole deformations  $\beta_4$, $\delta_4$,
and $\gamma_4$ [Eq.~(\protect\ref{q4def})] as functions of
 angular momentum for $j$=19/2,  $\chi_4$=0.5, and $N$=8
(left) and 10 (right).
Values of $\tilde\delta_4$ and $\tilde\gamma_4$ obtained in the
strong-coupling limit [Eq.~(\protect\ref{roh1})] are denoted by
dashed lines. The limits of axial hexadecapole shapes are
indicated by dotted lines ($\delta_4^{(x)}$, $\gamma_{4}^{(x)}$:
axial symmetry
with respect to $x$-axis; $\delta_4^{(z)}$, $\gamma_{4}^{(z)}$:
axial symmetry with respect to $z$-axis).
}
\end{figure}

\begin{figure}
\caption{\label{b4}
Equilibrium hexadecapole deformation $\beta_4$ as a function of
 $\chi_4$ for the HF
yrast states at a fixed angular momentum, $I$=40.
Calculations were performed for
 the
$j$=19/2 shell filled with $N$=8 (solid line) or $N$=10
(dashed line) particles.
}
\end{figure}

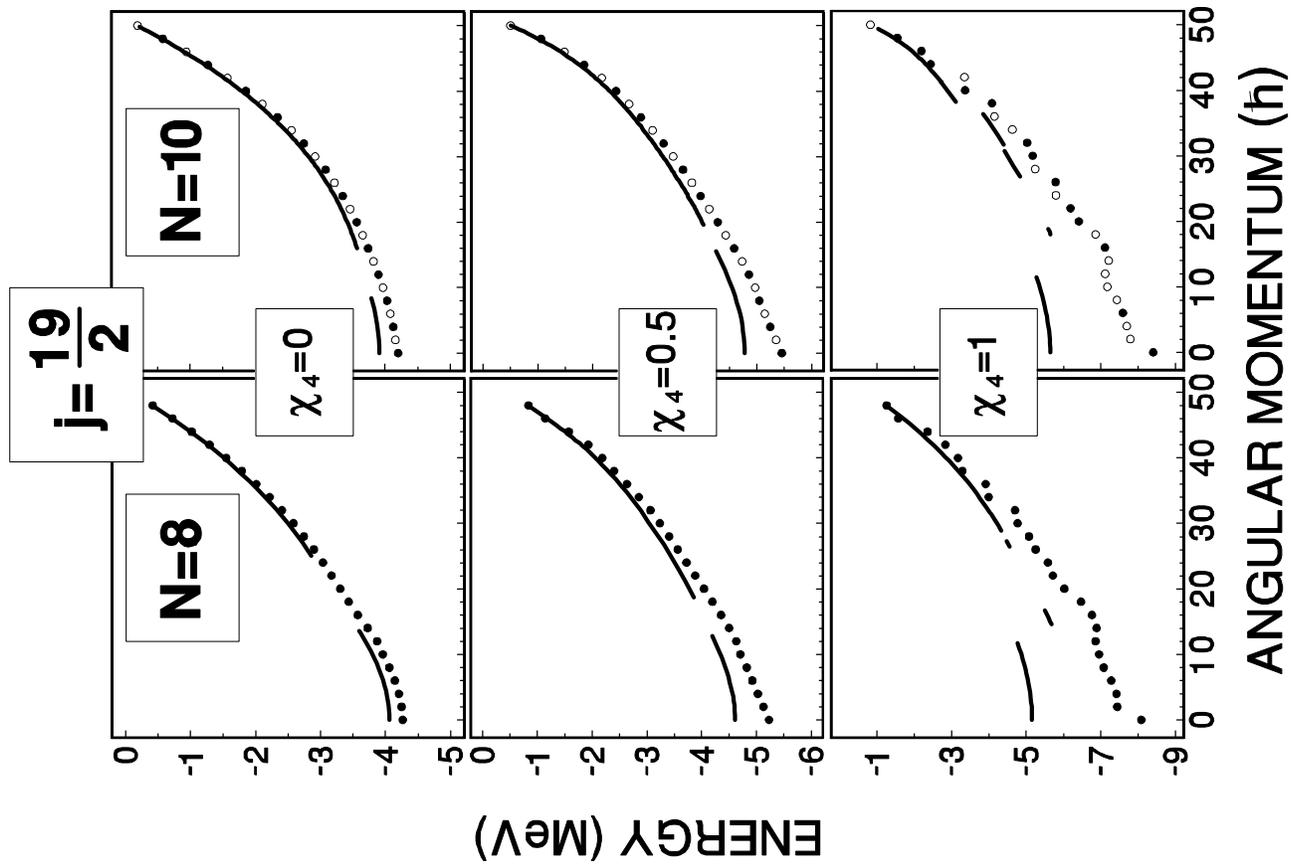
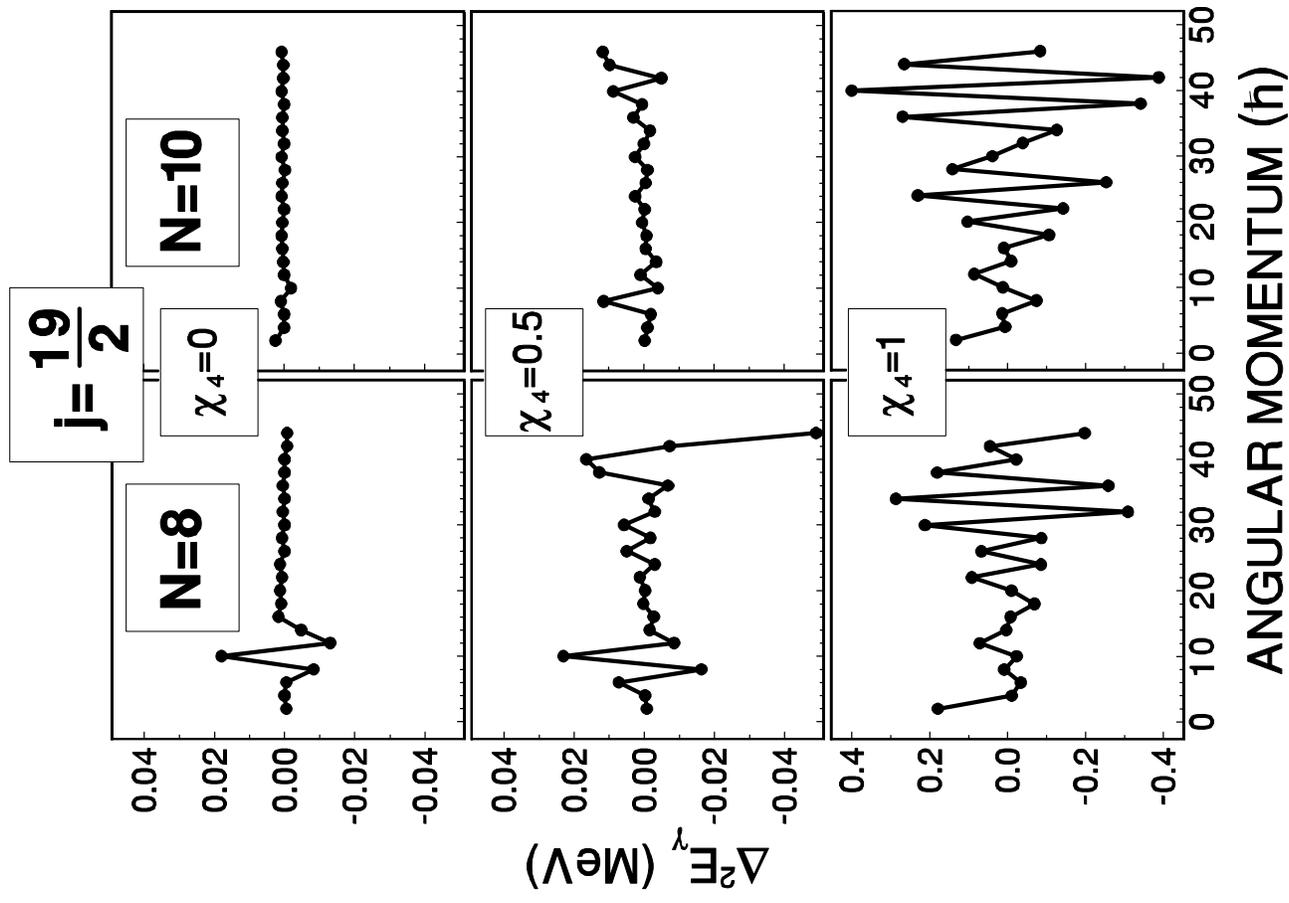
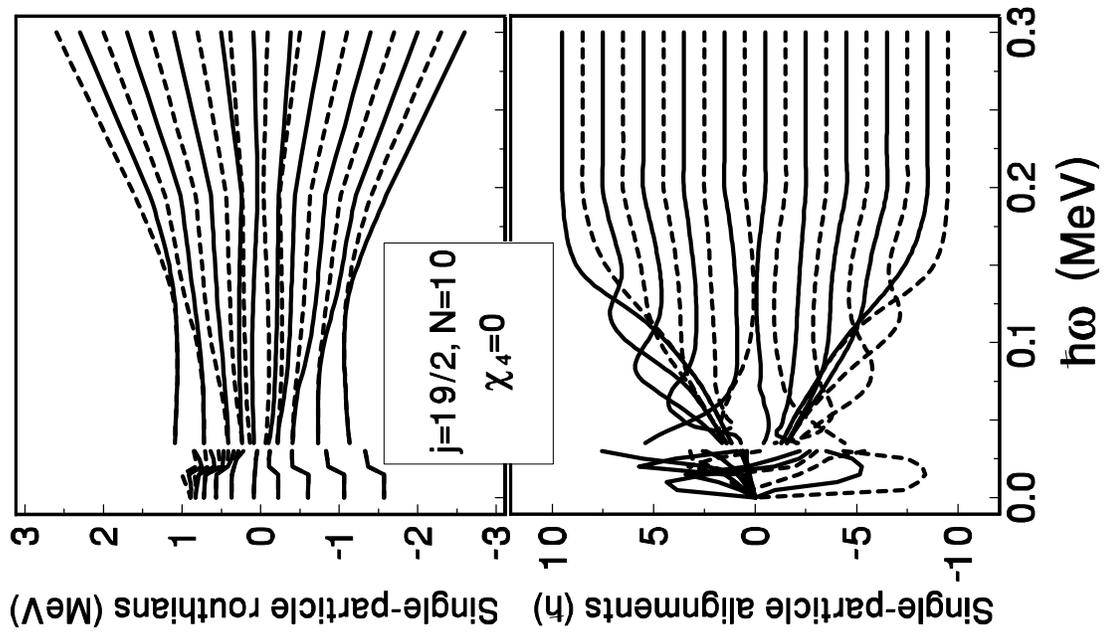
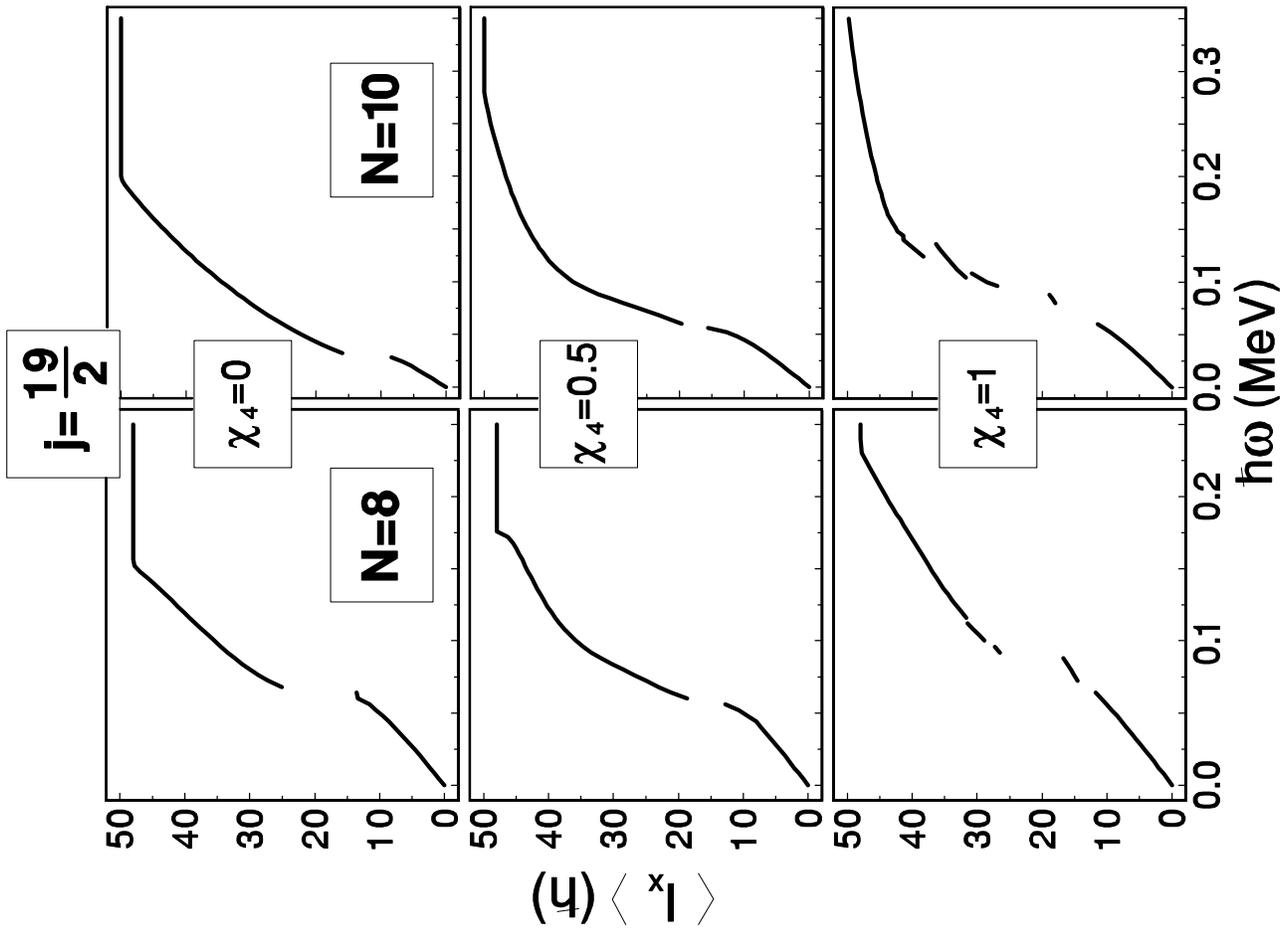
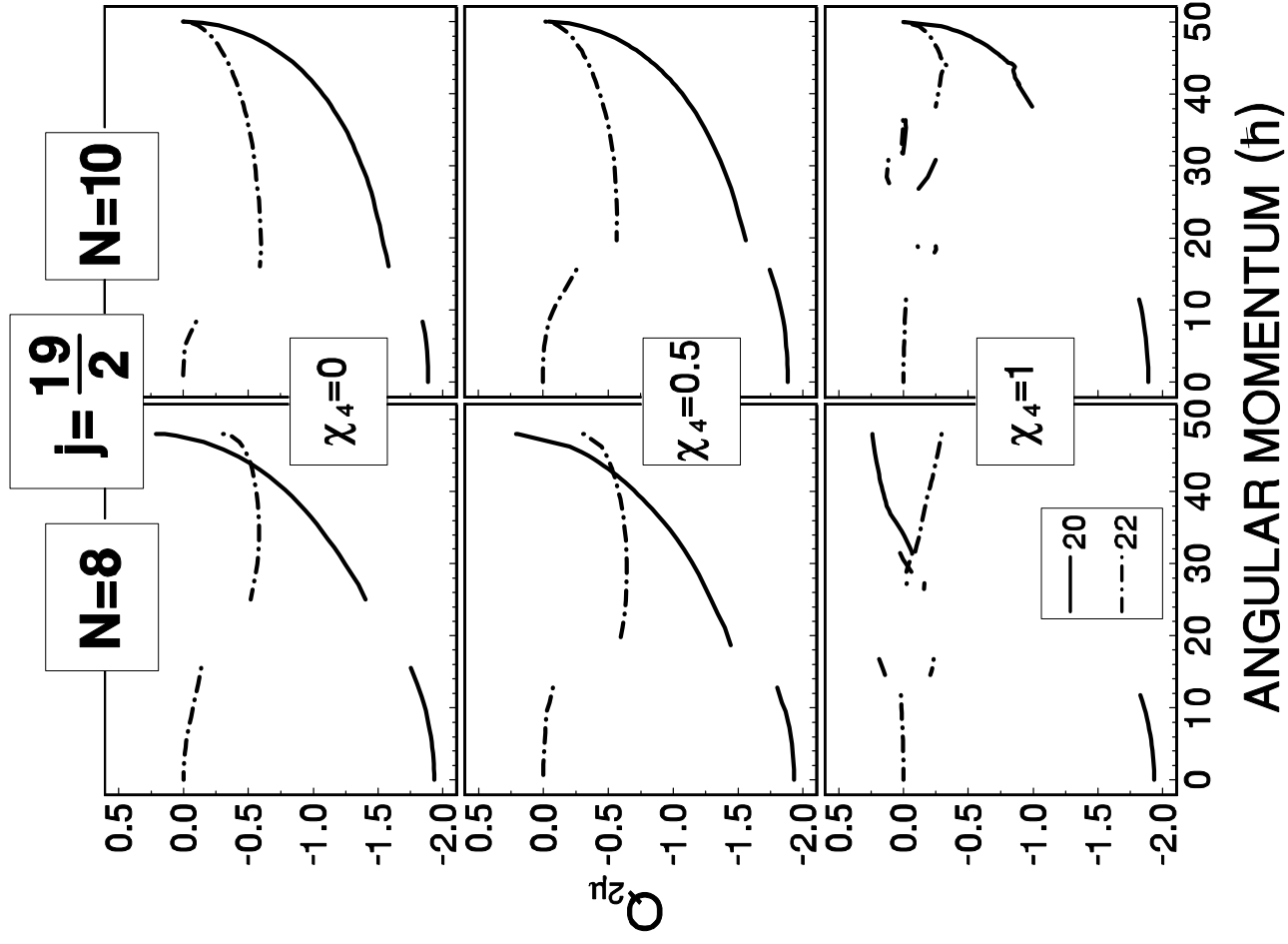
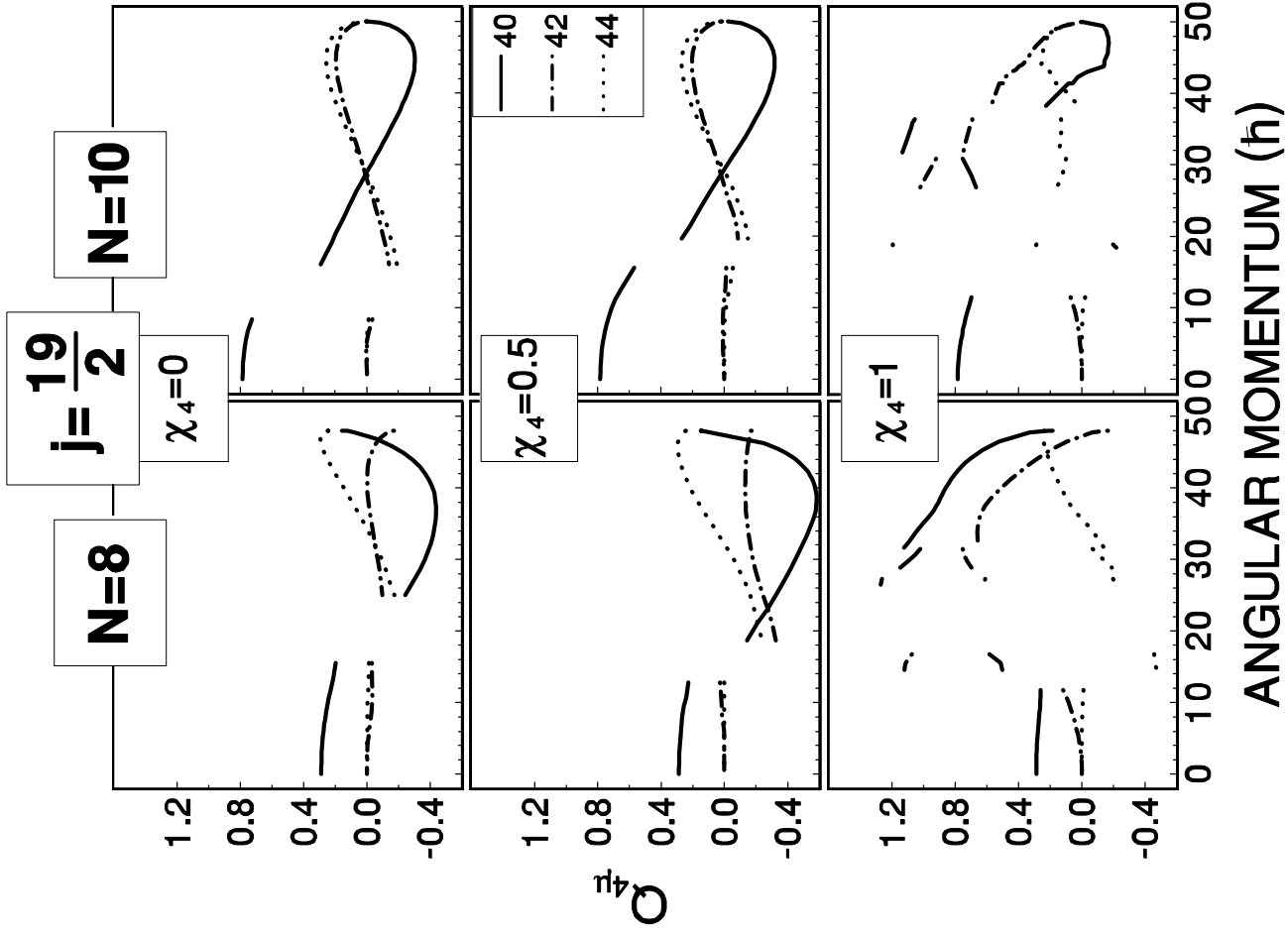
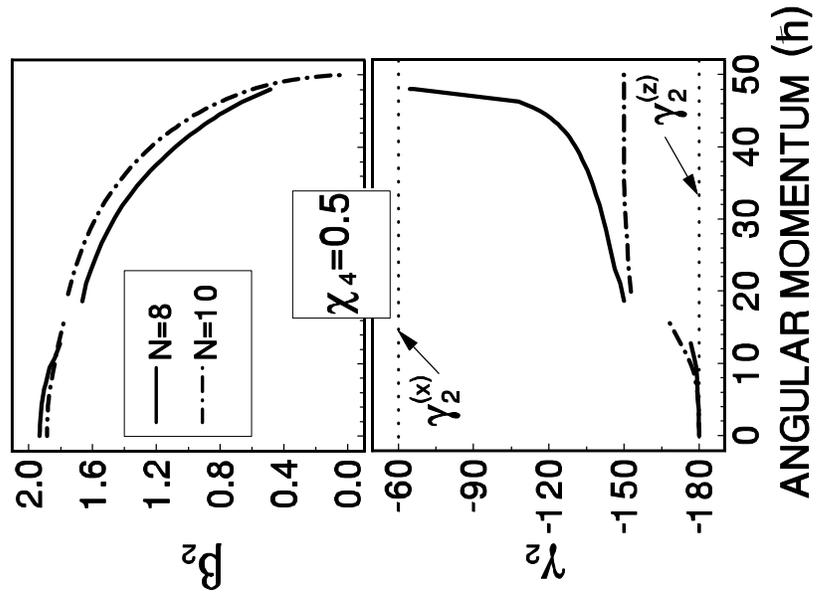
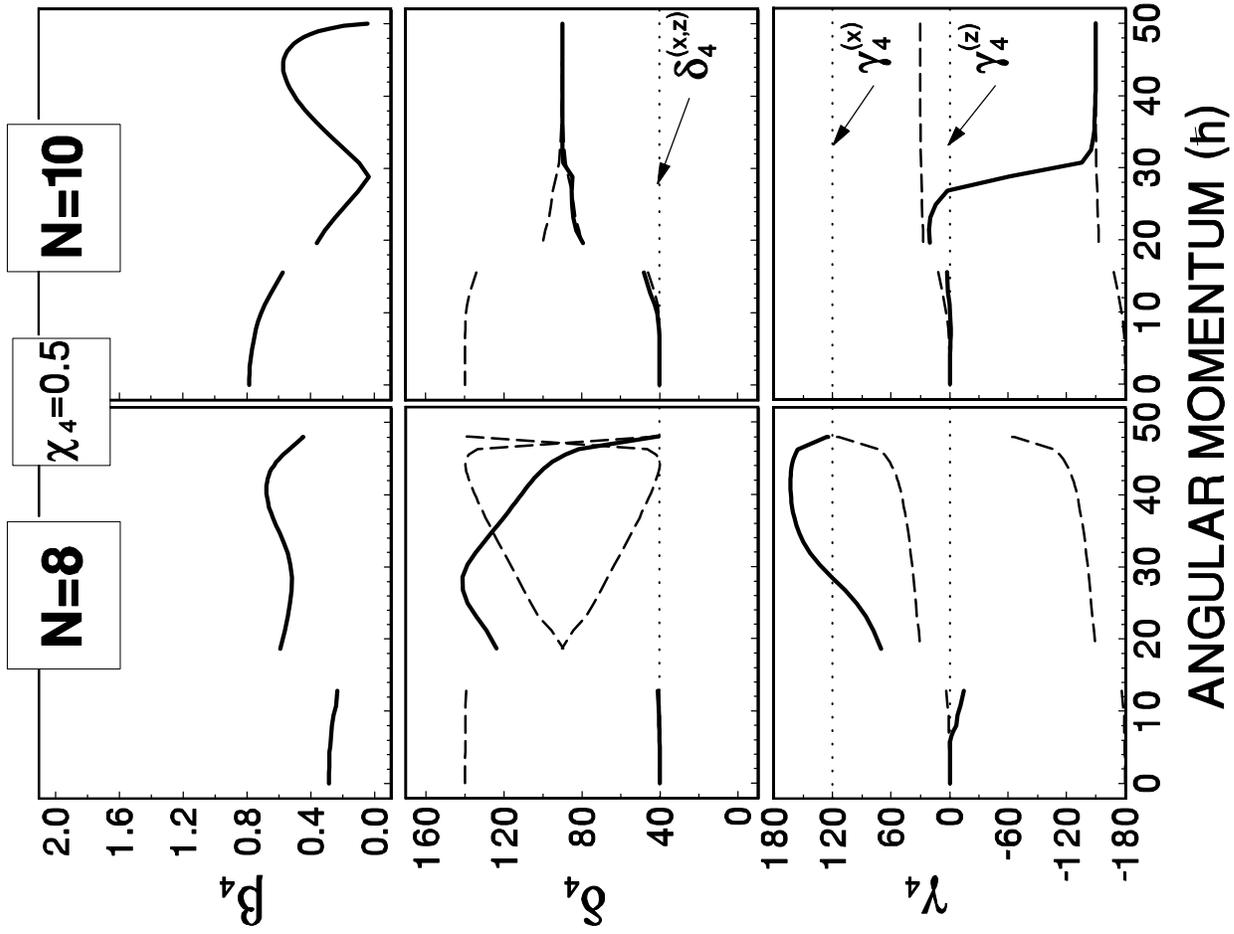
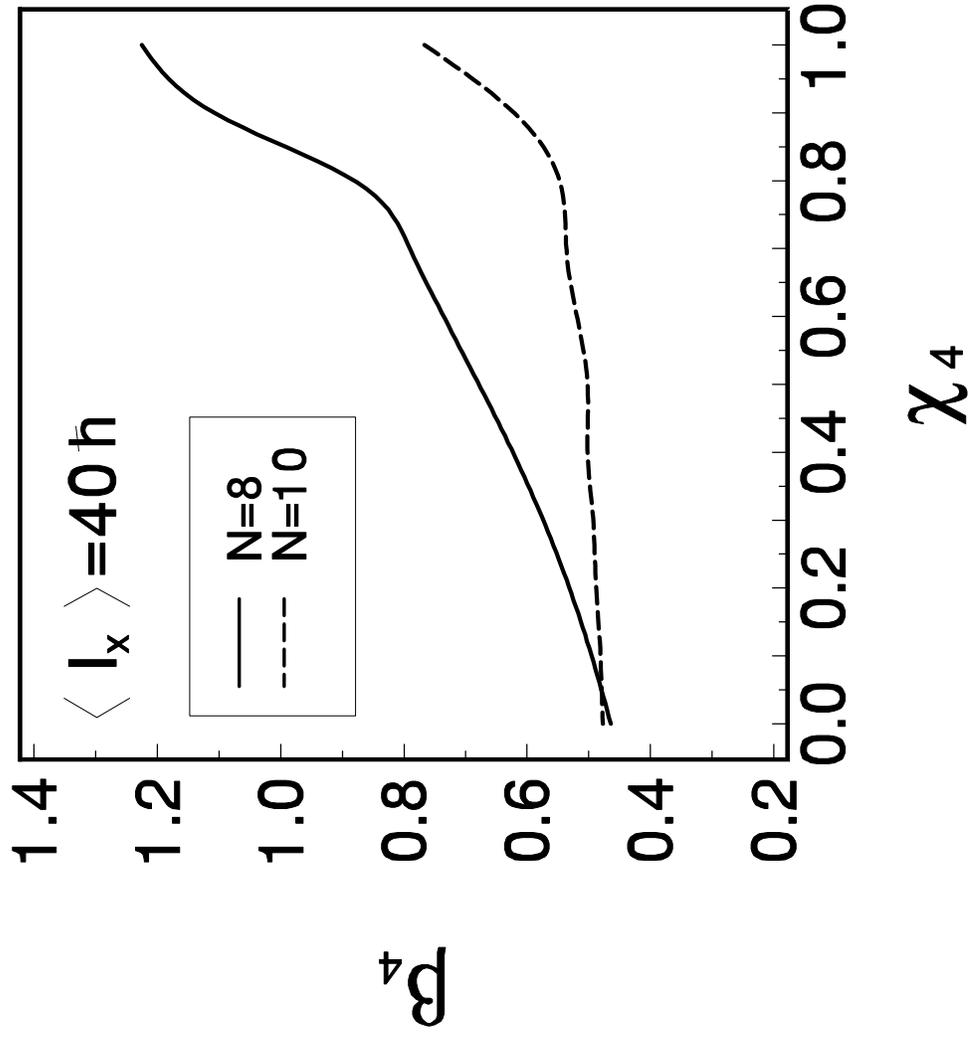
\end{document}